\documentclass[12pt]{article}

\usepackage{mathtools}
\usepackage{amssymb}
\usepackage{caption}
\usepackage{enumitem}
\usepackage[colorlinks=true,allcolors=blue]{hyperref}
\usepackage[sorting=none, backend=biber, url=false, isbn=false, style=nature]{biblatex}

\usepackage{xcolor}
\definecolor{cmmt}{rgb}{0.0, 0.5, 0.0}
\usepackage{bm}
\usepackage{courier}
\usepackage{times}
\usepackage{setspace}
\usepackage{threeparttable}
\usepackage{footnote}

\makesavenoteenv{tabular}
\makesavenoteenv{table}

\usepackage{algorithm,algorithmicx,algcompatible}
\usepackage[noend]{algpseudocode}

\usepackage{siunitx}
\settoggle{bbx:url}{false}
\AtEveryBibitem{\clearfield{issn}}
\AtEveryCitekey{\clearfield{issn}}
\AtEveryBibitem{\clearfield{month}}
\AtEveryCitekey{\clearfield{month}}

\title{\huge Scalable multicomponent spectral analysis\\ for high-throughput data annotation}

\author
{R. Patrick Xian$^{1,2\ast}$ \quad Ralph Ernstorfer$^{1}$ \quad Philipp M. Pelz$^{3,4}$\\
\\
\normalsize{$^{1}$Fritz Haber Institute of the Max Planck Society, 14195 Berlin, Germany.}\\
\normalsize{$^{2}$Department of Neurobiology, Northwestern University, Evanston 60208, IL, USA.}\\
\normalsize{$^{3}$National Center for Electron Microscopy Facility, Molecular Foundry,}\\
\normalsize{Lawrence Berkeley National Laboratory, Berkeley, CA 94720, USA.}\\
\normalsize{$^{4}$Department of Materials Science and Engineering,}\\
\normalsize{University of California, Berkeley, CA 94720, USA.}\\
\normalsize{$^\ast$Correspondence authors:  xian@fhi-berlin.mpg.de}\\
}

\date{}

\begin{document} 


\baselineskip20pt

\maketitle

\begin{quote}
\noindent\textbf{Orchestrating parametric fitting of multicomponent spectra at scale is an essential yet underappreciated task in high-throughput quantification of materials and chemical composition. To automate the annotation process for spectroscopic and diffraction data collected in counts of hundreds to thousands, we present a systematic approach compatible with high-performance computing infrastructures using the MapReduce model and task-based parallelization. We implement the approach in software and demonstrate linear computational scaling with respect to spectral components using multidimensional experimental materials characterization datasets from photoemission spectroscopy and powder electron diffraction as benchmarks. Our approach enables efficient generation of high-quality data annotation and online spectral analysis and is applicable to a variety of analytical techniques in materials science and chemistry as a building block for closed-loop experimental systems.}
\end{quote}

\section*{I. \, Introduction}
Real-time understanding of experimental data in materials and chemical characterization is a pursuit for the coming age of automation \cite{Spurgeon2020}. Building machine learning models for experimental techniques that produce spectral or spectrum-like data requires tools for \textit{spectrum annotation}, which pairs data with quantitative information extracted by fitting approximate lineshape or structural models \cite{Woodruff2016}. Achieving computational efficiency in these tasks allows processing large amounts of data with negligible human intervention as well as execution of online or on-the-fly analysis \cite{Kusne2020,Attia2020}. In materials characterization measurements, changes in sample and environmental conditions could appear in minutes to hours, while experimental campaigns often run continuously for days or longer. Moreover, the data acquisition process often involves tuning of external variables \cite{Hofmann2020} such as the spatial locations, time, temperature, concentration, etc., resulting in multidimensional datasets of multicomponent spectra (each component being a spectral lineshape or background model). As a realistic example, fitting one thousand 10-component spectra, with 3-4 parameters for each component would result in a large-scale optimization problem involving on the order of $10^4$ parameters. To assess the experimental outcome on this scale is not uncommon in modern high-throughput experimentation \cite{Umehara2019}. The dataset shifts during the experiment also make it challenging to apply modern machine learning frameworks without sufficient human interaction \cite{Kusne2020,Attia2020}. Therefore, data annotation methods with efficient computational scaling with respect to spectral complexity, quantified by the number of components, becomes crucial for automating materials characterization.

In this work, we leverage high-performance computing (HPC) to achieve linear scaling for a class of spectrum annotation task -- multicomponent spectral analysis (MSA) \cite{Blackburn1965} or fitting. Broadly speaking, the MSA problem encompasses two complementary scenarios: (1) Fitting an unknown number of peaks at fixed spectral locations with varying amplitudes; (2) Fitting a fixed number of peaks at varying spectral locations. Ultimately, MSA extracts information such as the shape and height of spectral components and disentangles signal from background. Mathematically, MSA may be mapped to separable nonlinear least squares problems \cite{Ruhe1980} present in various branches of science and engineering. In analytical chemistry, where scenario (1) is frequently encountered for understanding spectroscopic data, matrix factorization techniques without \cite{DeJuan2014} or with limited prior knowledge \cite{Kriesten2008} of pure spectral components (i.e. spectra from single sources or species) are often adopted for data analysis. In comparison, scenario (2) remains a recurring challenge in upscaling the analysis pipeline of quantitative electron- and X ray-based characterization techniques, when large, annotated, empirical databases are yet to exist for training machine learning models. Moreover, the equivalents in scenario (2) of pure spectral components are generally unattainable. Currently, relevant softwares for materials characterization usually only provide the option for sequentially fitting individual spectra \cite{Speranza2019} or pattern in the case of diffraction \cite{RenedeCotret2018}. Therefore, we have developed \texttt{pesfit} \cite{Xian}, a software to tackle the MSA problem in (2) at scale to facilitate online and offline analysis and experimental feedback. In the following, we discuss the software architecture, validate the computational scaling on batch (or distributed) spectrum fitting and present use cases in materials characterization and discuss the insights gained from data.

\section*{II. \, Results}
\subsection*{I. \, Scalable software architecture}
Fitting a multicomponent spectrum requires iterative calculation of its components, which are typically modelled as unimodal probability distributions or analytical functions. The mathematical context is provided in Appendix B. In practice, a complex spectrum may comprise tens of overlapping components or may be divided into segments of a similar size. Maintaining a reasonable computational performance with a large number of spectra with increasing complexities requires both resource and architectural upgrade on almost all steps of single-spectrum fitting that remains the mainstay in existing data analysis workflows \cite{Speranza2019,RenedeCotret2018}.
\begin{figure}[htb!]
  \begin{center}
    \includegraphics[scale=0.32]{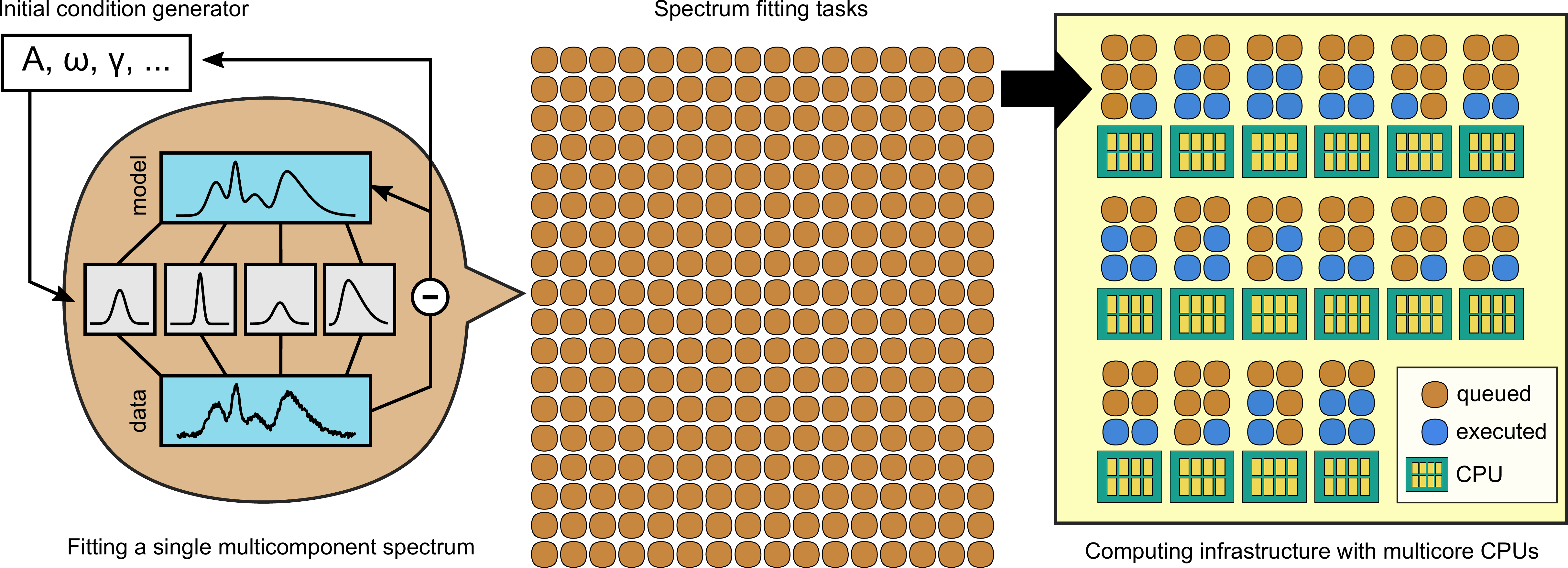}
    \caption{\textbf{Schematic of the scalable software architecture.} Each fitting task for a multicomponent spectrum (left) requires iterative evaluation of an approximate spectral lineshape model. A MapReduce approach implemented in \texttt{MultipeakModel} (see Appendix B) is used to execute the calculation of all components. During optimization, the fitting parameters are updated to refine the outcome, guided by a least squares error function involving the data and the fit. The entire pool of fitting tasks (middle) formed by aggregating individual ones is distributed over the multicore CPUs in a high-performance computing infrastructure (right) and different task schedulers may be used. The cartoon shows 16 worker processes each assigned with 6 fitting tasks. Task-based parallelization with the asynchronous scheme shown here (right) allows to balance the computational loads between fitting tasks.}
    \label{fig:architecture}
  \end{center}
\end{figure}
We have realized linear scaling in \texttt{pesfit} to satisfy the computational needs and alleviate the manual task load in (1) multicomponent spectrum computation, (2) batch execution, as well as the associated steps of (3) initial condition generation and (4) fitting result collection. To minimize redundancy in (1), we constructed a generating class (\texttt{MultipeakModel}, see Appendix B) to represent multicompoent spectra, making use of a MapReduce structure \cite{Lammel2008} for evaluating the spectra given the constituents (see Fig. \ref{fig:architecture} left). Our approach extends the popular spectral lineshape fitting framework \texttt{lmfit} \cite{Newville2019} to allow batch fitting a multitude of spectra with an arbitrary number of components. The MapReduce approach splits up (``map" step) the evaluation of multicomponent spectra into parallelizable computation of the components followed by merging (``reduce" step) of the results (see Fig. \ref{fig:architecture} left) and allows to generate an arbitrary complex spectral shape from existing lineshape models or user-defined functions without hard-coding each type of multicomponent spectrum separately. Each task thus constructed executes fitting of a single spectrum (see Eq. \eqref{eq:SNNLS} in Appendix B). To enable (2), the compartmentalized fitting tasks (see Fig. \ref{fig:architecture} middle) are executed concurrently through \texttt{dask} \cite{DaskDevelopmentTeam2016}, \texttt{parmap} \cite{Oller}, and \texttt{torcpy} \cite{Hadjidoukas2020} on an HPC infrastructure with multicore CPUs (see Fig. \ref{fig:architecture} right). Resource sharing (including the numerical data and the multicomponent spectrum model) between tasks is implemented in distributed fitting (\texttt{DistributedFitter}, see Appendix B) to reduce the memory footprint. To simplify (3), the \texttt{pesfit} package includes convenient methods to programmatically generate initial conditions for a large number of fitting parameters (see Appendix B). The initial conditions are assembled into a nested key-value pairs (Python dictionary) to deliver to the optimizer. In step (4), the fitting results in the form of parameter lists are collected by a DataFrame data structure from \texttt{pandas}, whose expansive functionality allows further statistical analysis of the outcome \cite{mckinney-proc-scipy-2010}. The fitting results and the initial conditions may be exported for reproducible examination of the specific fitting tasks. A schematic of the computational workflow is shown in SM Fig. \ref{fig:workflow}.

\subsection*{B. \, Performance evaluation with use cases}
We demonstrate the use of batch fitting on experimental data from two popular materials characterization techniques: photoemission spectroscopy and powder electron diffraction. Both of these techniques generate multidimensional datasets involving multicomponent spectra or spectrum-like signals. Spectrum fitting is carried out with predefined, domain-specific multicomponent models (see Appendix C).
\begin{figure}[htb!]
  \begin{center}
    \includegraphics[width=\textwidth]{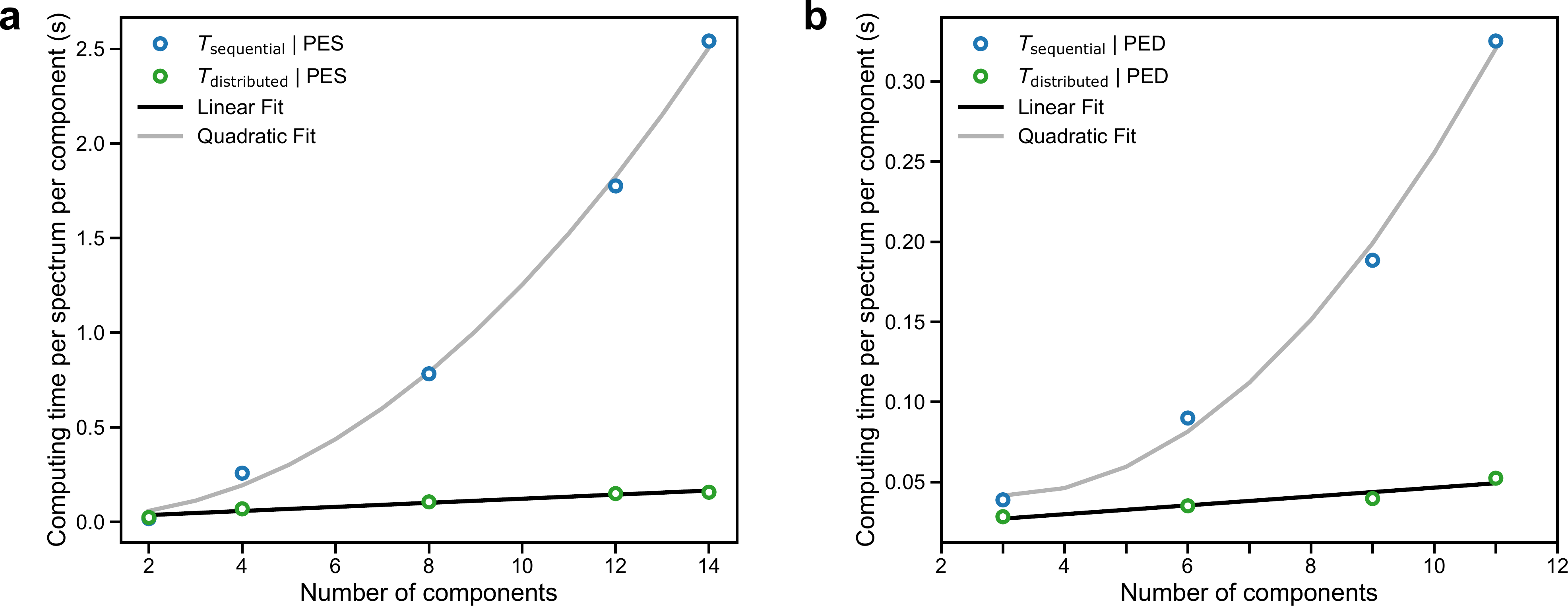}
    \caption{\textbf{Computational evaluation on experimental datasets.} Performance comparison between the sequential fitting procedure and the distributed fitting procedure on benchmark datasets: \textbf{a}, photoemission spectroscopy (PES) of the valence band (around K point in momentum space) of WSe$_2$ with up to 14 energy bands considered; \textbf{b}, Time-resolved powder electron diffraction (PED) of polycrystalline Pt \cite{Zahn2020}, with up to 10 diffraction peaks and a background profile considered. The computing times ($T_{\mathrm{sequential}}$ and $T_{\mathrm{distributed}}$), here used as the evaluation metric, are normalized by both the number of components and spectra (or diffraction peaks).}
    \label{fig:benchmarks}
  \end{center}
\end{figure}
In momentum-resolved photoemission spectroscopy (PES), extracting electronic band dispersion from the 3D dataset requires batch fitting of momentum-dependent intensity-valued energy spectra. We use photoemission band mapping data of bulk tungsten diselenide (WSe$_2$), which was measured for about an hour and contains a maximum of 14 energy bands within the probed energy range of $\sim$ 8 eV, to benchmark the fitting algorithms. The electron diffraction dataset was recorded for polycrystalline platinum (Pt) with a pump-probe scheme -- pump light induces electronic and lattice dynamics that are probed by electron diffraction -- to obtain time- and fluence-dependent 1D diffraction patterns \cite{Zahn2020}, resulting in 3D dataset. The data was acquired continuously over 2 days with up to 10 diffraction peaks of Pt within the measured reciprocal space. Further details of the experiments are provided in Appendix A.

In both use cases, the batch fitting is initiated with realistic guesses in their respective domains (see Appendix C), the optimization processes using both sequential and distributed approaches are timed for comparison. We compare the computing time (see Fig. \ref{fig:benchmarks}) that estimates the duration between the start of the fitting until the results are collected, which accounts for the accumulated wall time of steps (1), (2), and (4) mentioned earlier. The computing times are normalized to the per-spectrum and per-component level to allow comparison between datasets from different experimental modalities. As shown in Fig. \ref{fig:benchmarks}, within the tested cases, the distributed approach generally exhibits a linear dependence on the number of components in the data, while the sequential approach, as is common practice in the field, has at least a quadratic scaling. The disparity in normalized computing time between the two use cases reflects characteristics of the fitting tasks (spectrum model, initialization, hyperparameters, etc). Nevertheless, in both cases, distributed fitting of a significant portion of the data is sufficiently faster than the respective experimental duration, further discussed in Appendix C. The polynomial speed-up therefore allows multicomponent spectral analysis on the fly during the experimental measurements, with only the added cost on the computing hardware. Moreover, the tuning of the initialization to improve the fitting outcome may be conducted interactively using the same software, which allows to design custom approaches to deal with complex spectra. Some intuitive approaches to achieve high performance and consistency within the fitting results are discussed in the SM and schematized in SM Fig. \ref{fig:workflow}.

Specifically for the PES data, SM Fig. \ref{fig:finetuning} shows that the batch fitting method provides rapid access to the local dispersion in an extended region ($\sim$ 0.6 {\AA} $\times$ 0.6 {\AA} covered by 1600 energy spectra) around K point of the surface Brillouin zone of WSe$_2$, therefore uncovers the different strength of the trigonal warping (TW) feature in the energy bands visible by eye. TW is a third-order angular modulation \cite{Kormanyos2015} of the ideal parabolic dispersion (representing isotropic dispersion without angular distortion) often depicted for semiconductors. It is a manifestation of the strong spin-orbit interaction within the material \cite{Kormanyos2013} and has influence over the transport properties \cite{Yu2014,Shan2015,Kalameitsev2019}. Although theoretical descriptions exist \cite{Kormanyos2013,Kormanyos2015}, so far it has not yet been densely reconstructed directly from experimental data. The much improved throughput of our method now allow us to further study the details of TW in other energy bands of WSe$_2$ and compare with theoretical calculations.

\section*{III. \, Discussion and conclusion}
We have described a linear-scaling software architecture for mapping the multicomponent spectral analysis problem to distributed hardware. Our approach fully exploits the HPC infrastructure to provide a convenient toolkit for speeding up data annotation in materials characterization without requiring repeated data loading and human intervention. In the use case demonstrations, although we have assumed prior knowledge for the number of components, the requirement may be achieved using numerical estimation \cite{Lindner2015,Ament2019}. The modular architecture of the associated software, \texttt{pesfit}, allows convenient upgrade of its components individually in the future. For example, it is possible to plug in alternative optimizers for batch fitting, such as brute-force search or variable projection \cite{Golub2003}, which have respective speed-accuracy trade-offs in solving large-scale nonlinear least squares problems. The computational performance demonstrated here provides a baseline for future algorithm development using such annotations for training supervised learning algorithms. Besides the use cases discussed here, we envision the method and software to be applicable to techniques such as transient vibrational spectroscopy \cite{Grubb2016}, X-ray fluorescence imaging \cite{Pushie2014}, to name a few, where spectral shift along with the changes in spectral width and amplitude have unique physical and chemical implications, and where multicomponent spectral analysis are adopted for analyzing the measured multidimensional datasets. The software may be integrated with existing computational workflows \cite{RenedeCotret2018,Xian2020}, thereby functioning as a building block for closed-loop materials or chemical characterization and analytics \cite{Kusne2020,Attia2020} to facilitate experimental optimization and discovery.

\subsection*{Appendix A: \, Materials characterization datasets}
Both experimental datasets are measured with home-built instruments at the Fritz Haber Institute. The photoemission spectroscopy experiment is conducted using pulsed light source based on high harmonic generation as described previously \cite{Puppin2019}. The sample under study is single crystalline tungsten diselenide (WSe$_2$). Extreme UV pulses with an energy of 21.7 eV liberate electrons from the WSe$_2$ sample surface, which are subsequently captured as single events in 3D (energy and two momenta) by a time-of-flight delay-line detector (METIS 1000, SPECS GmbH). A custom computational pipeline is used to preprocess the single events to produce the 3D photoemission band mapping data \cite{Xian2020}. The time-resolved powder electron diffraction experiment \cite{Zahn2020} contains fluence-dependent measurements of polycrystalline Pt diffraction pattern with the laser excitation at 0.70 eV and 70 kV electron bunches generated from a gold photocathode \cite{Waldecker2015}. The 2D diffraction patterns are recorded with a CMOS camera (TemCam-F416, TVIPS GmbH) at a frame integration time of 5 s. The measured diffraction rings are collapsed into 1D diffraction peaks via radial averaging.

\subsection*{Appendix B: \, Software details and mathematical context}
The main software modules in \texttt{pesfit} are \texttt{lineshape}, \texttt{fitter} and \texttt{metrics}. The \texttt{lineshape} module contains the \texttt{MultipeakModel} class. The \texttt{fitter} module contains funtions to automate the construction of fitting tasks and initializations as well as simple visualization methods for displaying fitting results. The \texttt{metrics} module includes evaluation metrics for assessing the quality of fits. The sequential and distributed fitting are achieved through the \texttt{PatchFitter} and \texttt{DistributedFitter} classes, respectively, both lie within the \texttt{fitter} module. To fit higher-dimensional ($>$ 2D) spectral data, the data are first partially transformed to 2D before the fitting tasks are farmed out to processors. The initial conditions can be generated programmatically using \texttt{init\_generator} and supplied to a multicomponent model via \texttt{varsetter}, both are integrated into the \texttt{fitter} module classes discussed here, but are also invocable separately and individually. The ordering of fitting tasks is tracked by a spectrum index (\texttt{spec\_id}) to accommodate asynchronous concurrent execution, which permits index scrambling.

Both sequential and distributed fitting of the benchmark datasets uses the default Levenberg-Marquardt algorithm for minimization with a spectrum-wise least-squares loss function and bound constraints. We formulate the single multicomponent spectral analysis task as
\begin{equation}
    \Tilde{\textbf{S}} = \mathrm{argmin} \, \left\Vert I(\omega) - \sum_n f_{n}(\omega, \textbf{s}_n) \right\Vert_{\omega}^2.
    \label{eq:single_fit}
\end{equation}
Here, $I$ is the intensity-valued spectrum data with the spectral dimension $\omega$. $\textbf{s}_n = (s_{n,1}, s_{n,2}, ...)$ is the set of parameters of the single-component spectral lineshape or background model $f_n$, with $n$ being the index of spectral components. The symbol $\left\Vert \cdot \right\Vert_{\omega}$ denotes the Euclidean norm with respect to the spectral dimension. $\Tilde{\textbf{S}} = \left[ \Tilde{\textbf{s}}_1 \, \Tilde{\textbf{s}}_2 \, ... \right]$ is the concatenated list of component-wise fitting parameters after optimization. For example, a 10-component spectrum with 4 parameters for each component results in an $\Tilde{\textbf{S}}$ of size 40 for a single-spectrum fitting task. Based on this notation, batch fitting of multidimensional data is written as
\begin{equation}
    \{ \Tilde{\textbf{S}}^{\{a_i\}} \} = \mathrm{argmin} \, \left\Vert I(\omega, \{a_i\}) - \sum_n f_{n}^{\{a_i\}}(\omega, \textbf{s}_n^{\{a_i\}}) \right\Vert_{\omega}^2,
    \label{eq:SNNLS}
\end{equation}
where the spectrum indices $i$ are positive integers. Now, $I$ becomes multidimensional with the coordinates of other dimensions represented by $\{a_i\} = (a_1, a_2, ...)$. A scalar-valued spectrum index is assigned to each multidimensional coordinate $\{a_i\}$ according to the row-major order. On the left-hand side of Eq. \eqref{eq:SNNLS}, $\{ \Tilde{\textbf{S}}^{\{a_i\}} \} = \{ \left[ \Tilde{\textbf{s}}_1 \, \Tilde{\textbf{s}}_2 \, ... \right]^{\{a_i\}} \}$ involves a double concatenation (firstly row-wise over the inner brackets $\left[ \, \cdots \right]$ and ordered by the spectral components, secondly column-wise over the outer brackets $\{ \, \cdots \}$ and ordered by the spectrum index), which represents the entirety of spectrum-wise optimized parameter sets gathered over all $\{a_i\}$ coordinates. This arrangement turns the fitting results into a 2D array regardless of the dimensionality of the data. The nonlinear least-squares problems in Eqs. \eqref{eq:single_fit}-\eqref{eq:SNNLS} are considered separable in the sense that the spectrum model may be written as a sum of components \cite{Ruhe1980,Golub2003}.

\subsection*{Appendix C: \, Computational benchmarks}
All batch fitting benchmark use cases are run on an on-premises computing server (Dell PowerEdge R840), equipped with four Intel Xeon Gold 6150 multicore CPUs. The sequential fitting tasks are run with a single logical core, while the distributed fitting tasks are run using the designated number of processes mapped onto logical cores. Throughout the runs, the RAM size doesn't pose a limit on the computing time of the benchmark fitting tasks. Fitting of the photoemission data of WSe$_2$ uses the density functional theory electronic structure calculation available within the NOMAD repository (DOI: \href{http://dx.doi.org/10.17172/NOMAD/2020.03.28-1}{10.17172/NOMAD/2020.03.28-1}) along with a rigid global shift to all band energies as the initial condition for the band positions (in both sequential and distributed approaches), while fitting the diffraction peak positions of polycrystalline Pt uses the estimate positions from a reference spectrum \cite{Zahn2020}. The trends of the computational scaling of the two approaches persist regardless of the number of spectra used for benchmarking. For photoemission (see Fig. \ref{fig:benchmarks}a), the scaling was estimated by fitting 900 spectra. Each spectrum was fit with up to 14 Voigt lineshapes in combination. For experimental monitoring, band dispersion information from a region around a high-symmetry point of the material suffices. As a time reference, batch fitting of 1600 4-component photoemission spectra (see SM Fig. \ref{fig:finetuning}) costs around 6 mins.

The electron diffraction data was fit with up to 10 Voigt lineshapes and an exponential-decay background model. A total of 225 1D diffraction patterns were obtained by averaging from the over 4600 patterns measured in the 2-day experimental campaign. For computational scaling estimation (see Fig. \ref{fig:benchmarks}b), fitting the entire set of 225 11-component 1D diffraction patterns costs $<$ 3 mins. Since in practice, only a fraction of the data (both in range and in quantity) is needed to monitor the sample condition, the speed boost is sufficient to enable real-time interaction with experiments. Running the distributed fitting consistently at the optimal speed requires tuning of the task parallelization parameters -- the number of worker processes and the number of tasks per process. An extended example that illustrate the interdependence between these two parameters is shown in SM Fig. \ref{fig:taskparallel}. Nevertheless, in practice, only a few trials guided by heuristics on subsets of data are needed to find a reasonable trade-off that minimizes the computing time.


\subsection*{Acknowledgments}
We thank S. Dong, S. Beaulieu and L. Rettig for providing the photoemission spectroscopy data, D. Zahn for providing the electron diffraction data and helpful discussions. We thank S.~Sch{\"u}lke and G.~Schnapka at Gemeinsames Netzwerkzentrum (GNZ) and L. Rettig at the Fritz Haber Insitute in Berlin for support on the computing infrastructure. R.P.X. thanks S. Oller at Universitätsklinikum Hamburg-Eppendorf (UKE) for helpful discussion on code parallelization. R.P.X. and R.E. acknowledge the support by BiGmax, the Max Planck Society's Research Network on Big-Data-Driven Materials-Science, the European Research Council (ERC) under the European Union's Horizon 2020 research and innovation program (Grant No. ERC-2015-CoG-682843). P.M.P. acknowledges financial support from STROBE: A National Science Foundation Science \& Technology Center under Grant No. DMR 1548924.

\subsection*{Data Availability}
The data presented in the paper is available upon reasonable request from the corresponding author.

\subsection*{Author contributions}
R.P.X. conceived the project and developed the software with help from P.M.P.. R.P.X. wrote the first draft of the paper. All authors contributed to discussion and revision of the manuscript to its final form.

\subsection*{Competing interests}
The authors declare no competing interests in the content of the article.

\clearpage

\renewcommand{\thesection}{S\arabic{section}}

\renewcommand{\figurename}{SM Figure}
\renewcommand{\tablename}{SM Table}
\setcounter{figure}{0}

\renewenvironment{quote}
  {\small\list{}{\rightmargin=0.5cm \leftmargin=0.5cm}%
   \item\relax}
  {\endlist}

\begin{quote}
    \centering
    \huge Supplementary Material \\
    \vspace{0.5em}
    \huge Scalable multicomponent spectral analysis for high-throughput data annotation
\end{quote}
\vspace{2em}

\begin{centering}
\author
{\large R. Patrick Xian$^{1,2\ast}$ \quad Ralph Ernstorfer$^{1}$ \quad Philipp M. Pelz$^{3,4}$
\vspace{1em}
\\

\setstretch{1.2}
\normalsize{$^{1}$Fritz Haber Institute of the Max Planck Society, 14195 Berlin, Germany.}\\
\normalsize{$^{2}$Department of Neurobiology, Northwestern University, Evanston 60208, IL, USA.}\\
\normalsize{$^{3}$National Center for Electron Microscopy Facility, Molecular Foundry,}\\
\normalsize{Lawrence Berkeley National Laboratory, Berkeley, CA 94720, USA.}\\
\normalsize{$^{4}$Department of Materials Science and Engineering,}\\
\normalsize{University of California, Berkeley, CA 94720, USA.}\\
\normalsize{$^\ast$Correspondence authors:  xian@fhi-berlin.mpg.de}\\
}
\end{centering}
\vspace{5em}

\section{Modes of operation}
\begin{figure}[htbp!]
  \begin{center}
    \vspace{1em}
    \includegraphics[width=0.55\textwidth]{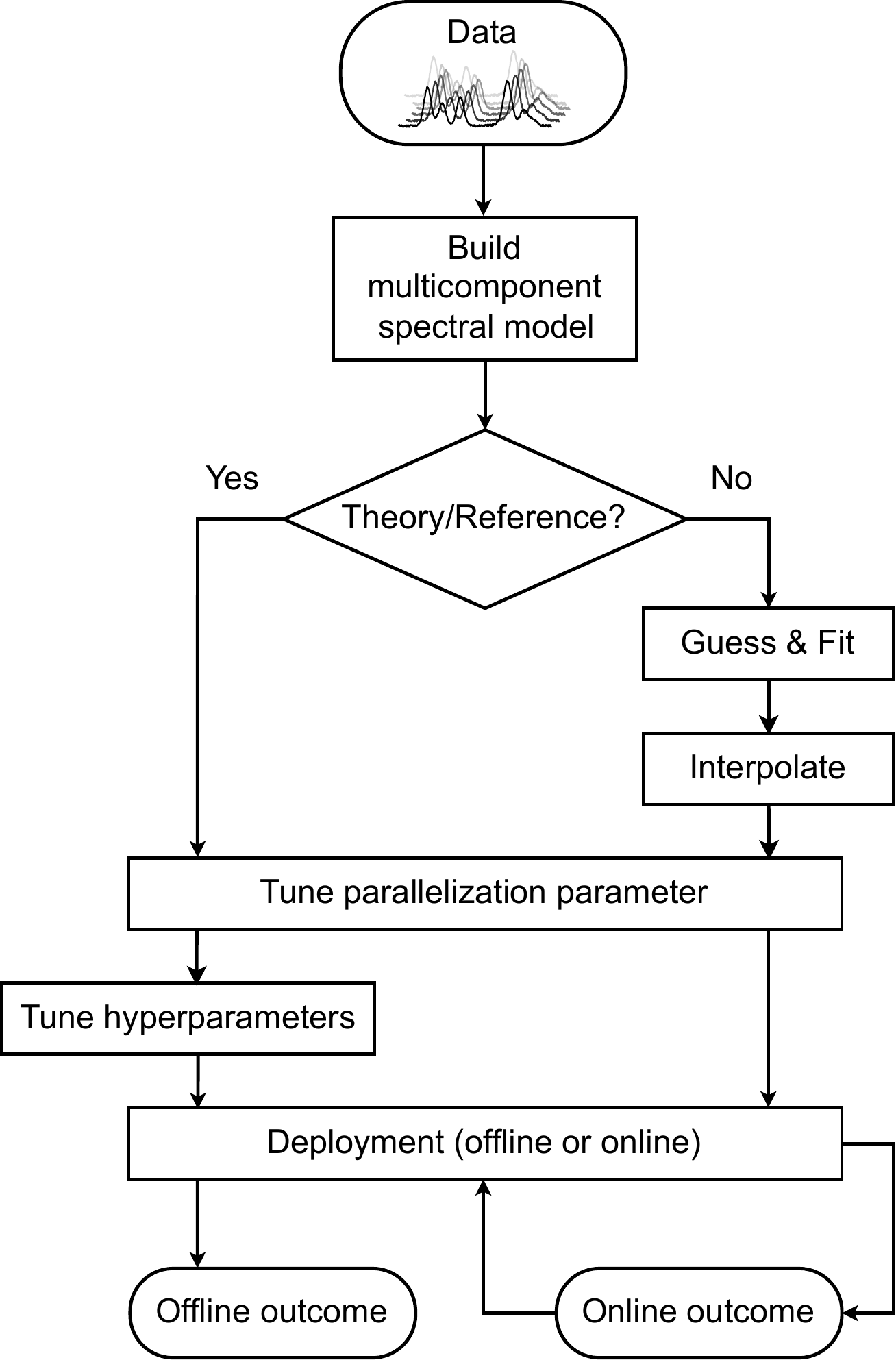}
    \vspace{1em}
    \caption{\textbf{Proposed multicomponent spectral analysis workflow.} The computational workflow starts from the dataset organized into a multidimensional array. Then, the multicomponent spectrum model is constructed using domain knowledge. If theoretical calculation or a reference spectrum is available, the workflow progresses to the stage of tuning task parallelization parameters. Otherwise, an initial guess fit of single spectra at distinct locations is required, along with interpolation of the spectral peak locations to other instances without guess fitting results. Using theory or reference spectrum positions as input, hyperparameters tuning is needed to compensate for the offsets from experimental data, but this is often unnecessary if guess fitting results are available. Finally, the batch fitting routine may be deployed for offline analysis or online monitoring of experimental results, using user-defined metrics constructed with the spectrum fitting outcome.}
    \label{fig:workflow}
  \end{center}
\end{figure}

Although the problem mapping and software realization presented in this work offer a scalable and automatable solution for batch fitting of multicomponent spectra, situations will always exist when a subset of the fitting tasks yield unfeasible answers due to experimental resolution, noise in the data, the sloppiness of the model used for fitting \cite{Transtrum2015}, etc. This also applies to offline data analysis where a more accurate model is often required for fitting to extract specific parameters, while in online analysis, approximate models with less computational overhead are favored. In the following, we offer here practical solutions outlined in the computational workflow (see SM Fig. \ref{fig:workflow}) to resolve potential outstanding cases while making use of the existing functionalities offered by \texttt{pesfit}. In the demonstrations below, we discuss primarily in the context of photoemission data due to its complexity compared with diffraction data, but a similar line of reasoning also applies to other materials characterization methods that produce spectra or spectrum-like data.

\subsection{Tuning of hyperparameters and initialization}
To arrive at a meaningful outcome, batch fitting of multicomponent spectra requires suitable initialization with knowledge from theoretical calculations or educated guesses. We treat rigid shifts applied to theory or reference in initialization as hyperparameters in the fitting procedure, which may be tuned according to the data characteristics. We discuss here three possible ways that may be realized using \texttt{pesfit}.
\vspace{1em}

\begin{figure}[htbp!]
  \begin{center}
    \includegraphics[scale=0.6]{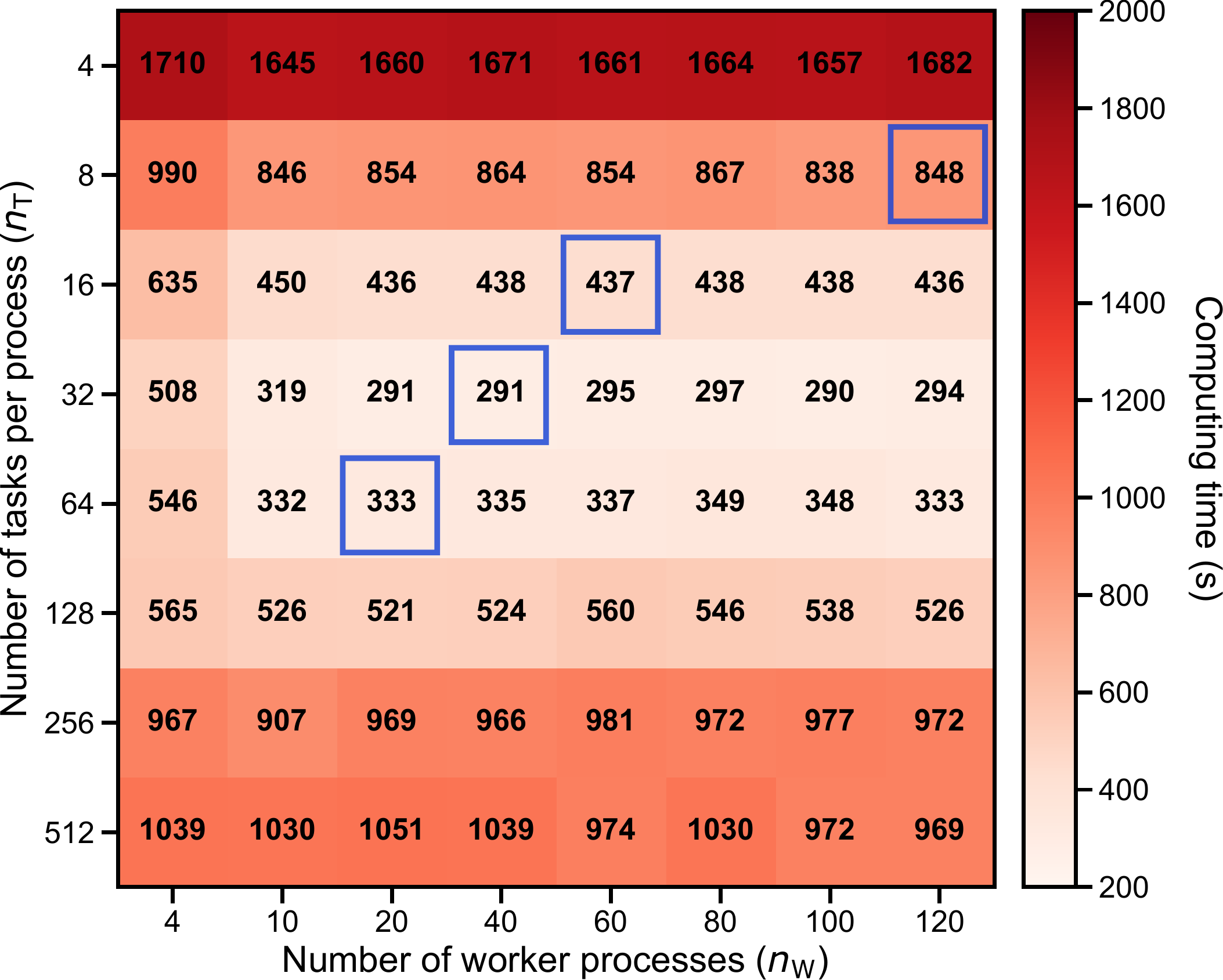}
    \caption{\textbf{Effects of task parallelization parameters on computing time.} Linear-scale heatmap of the computing time for the distributed multicomponent fitting procedure presented in this work. Each entry represents the computing time (in seconds) elapsed in a numerical experiment, where a distinct pair of parameters (processes and tasks per process) is selected to execute the same number of multicomponent spectral analysis tasks (900 spectra in total, each with 4 components). The benchmark uses the photoemission data from around the K point of WSe$_2$. The purple boxes represent the possible cases to try in the performance tuning guided by heuristics.}
    \label{fig:taskparallel}
  \end{center}
\end{figure}

\noindent\textbf{Theory-based approach}. When theoretical calculations or reference spectra are present, (i) The \texttt{random\_varshift} method in the \texttt{fitter} module of \texttt{pesfit} enables automated tuning of initial condition using random search within user-defined values to optimize a goodness-of-fit metric. (ii) Alternatively, if segments of the spectra may be isolated with no overlap with the rest (e.g. when spectral intensity goes to baseline level at the two ends of the segment), a (1) divide-and-conquer approach may be adopted to batch fit and tune the hyperparameters for a segment at a time. This applies to diffraction data and the photoemission data of materials with relatively flat and isolated bands, such as organic molecular crystals or monolayers \cite{Nakayama2020}. If the isolation between segments is less perfect,
\begin{figure}[htbp!]
  \begin{center}
    \includegraphics[width=0.85\textwidth]{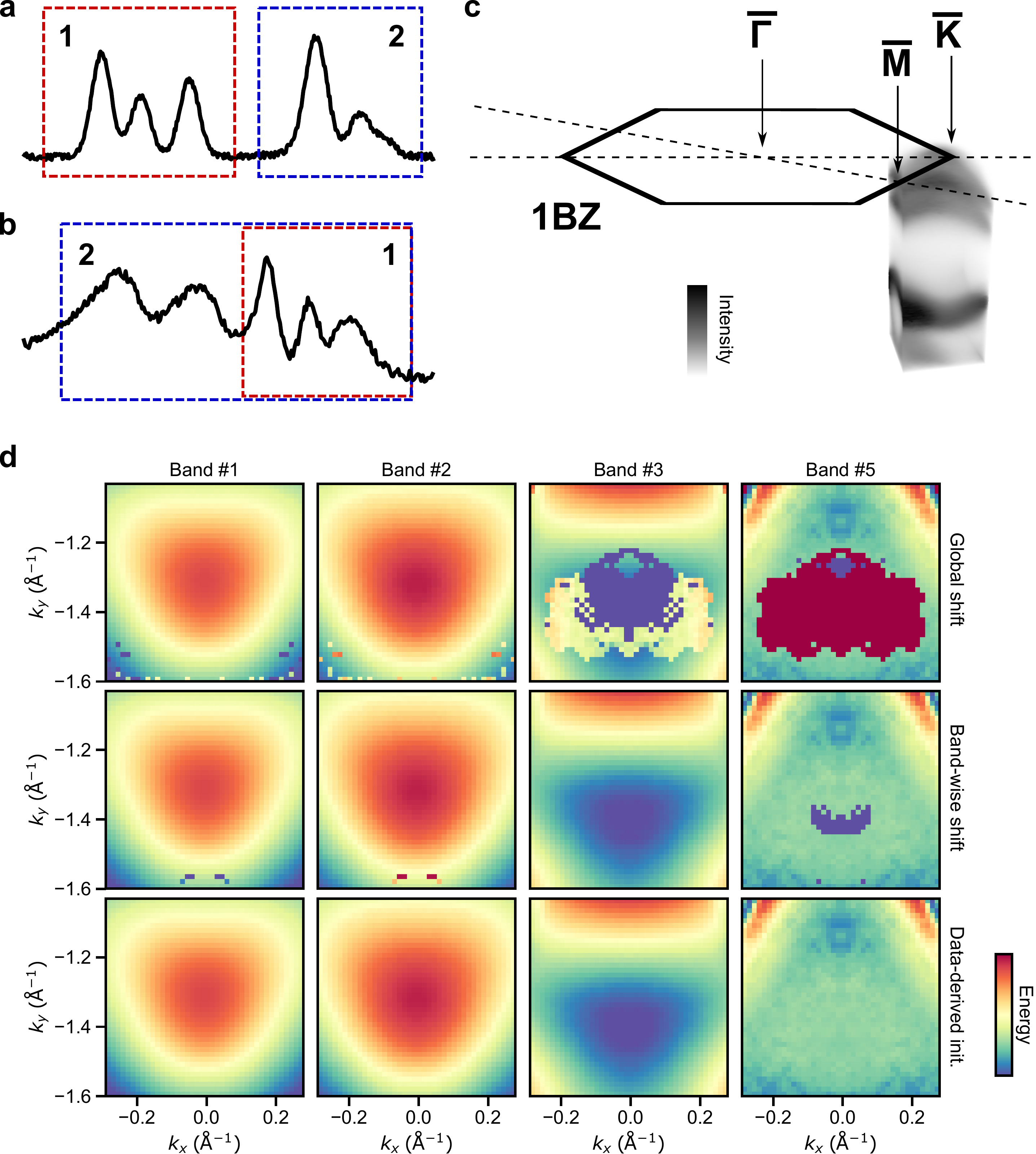}
    \caption{\textbf{Semi-automated fine-tuning of hyperparameters.} Tuning of the hyperparameters in batch fitting multicomponent spectra with \textbf{a}, the divide-and-conquer approach and \textbf{b}, the expanding window approach, illustrated with synthetic spectra. The two segments are numbered in dashed boxes with different colors. Demonstration of the tuning uses \textbf{c}. the photoemission dataset (containing 1600 multicomponent spectra) measured near the $\overline{\mathrm{K}}$ point of WSe$_2$. Other important landmarks within the projected first Brillouin zone (1BZ) include the zone center ($\overline{\Gamma}$) and zone edge ($\overline{\mathrm{K}}$). The overline signifies the surface projection (e.g. $\overline{\mathrm{K}}$ in photoemission corresponds to K). \textbf{d}. Comparison of the band dispersions reconstructed using batch fitting without (upper row) and with band-wise fine-tuning (middle row) of the shift hyperparameters of band structure calculation. The lower row shows the batch fitting results using data-derived initialization. The fourth band is not resolved here due to strong overlap with the third band at around K. The band mapping data used for fitting has been symmetrized to balance the matrix elements. The colormap for each energy band is scaled separately for visualization.}
    \label{fig:finetuning}
  \end{center}
\end{figure}
one may adopt an (2) expanding window approach to batch fit and tune an increasing number of spectral components (thus the ``expanding window"), retaining the hyperparameters used for the previous segments while fitting in an increasingly broader spectral window. These two approaches are illustrated in SM Fig. \ref{fig:taskparallel}a-b. During the tuning process, the fitting outcome may be appraised using either goodness-of-fit metrics or via visual inspection. An example showing the results before and after hyperparameter tuning is shown in SM Fig. \ref{fig:taskparallel}c-d using photoemission data from WSe$_2$. The results compared include batch fitting initialized with a 0.2 eV rigid shift of all bands (3c) and that initialized with a band-wise tuned rigid shift of 0.2 eV, 0.18 eV, 0.22 eV and -0.24 eV for each energy band (3d), respectively.
\vspace{1em}

\noindent\textbf{Data-driven approach}. When theoretical calculations or reference spectra are not available, (iii) one can resort to a data-driven approach. A selected number of distinctive spectra may be fit manually (guess fit) to determine the anchor points as reference for interpolation, which generates the initial conditions for the entire dataset. The data-derived initial conditions may then be used in batch fitting. An example using 16 equally-spaced anchor points to fit 1600 (40 $\times$ 40 grid in $k_x$ -- $k_y$ coordinates) multicomponent spectra is shown in SM Fig. \ref{fig:taskparallel}d (lower row). Here, we used the Clough-Tocher interpolator \cite{Alfeld1984} implemented in \texttt{scipy} \cite{Virtanen2020} to obtain the data-derived initial conditions for all bands, although other multivariate nonlinear interpolators should suffice in general. The approach is realized with the \texttt{InteractiveFitter} class within \texttt{pesfit}. In comparison, the data-driven approach shows higher consistency among neighboring spectra due to the closer resemblance of data-derived initial condition than the theoretical calculation to the final outcome.

\subsection{Tuning of model complexity}
In cases where strictly isolating a group of spectral features using data slicing is challenging, there may exist the situation where a subset of the data requires a more complex model (secondary model) to fit than the rest (primary model). Alternatively, the tuning may also be realized by changing the data range used in fitting. This appears commonly in cases where spectral shift is present, such as in photoemission data due to the existence of band dispersion and matrix element effects, which results in the dependence of spectral features on momentum coordinate (in momentum-resolved measurements), sample position (in spatially-resolved measurements), etc. Comparatively, in diffraction datasets (such as in \cite{Singh2017}) involving phase transition due to changes in macroscopic parameters or dynamics, the diffraction peak position changes, if any, are relatively small with respect to the peak width. After batch fitting with the primary model, the less well-fit spectra may be selected using goodness-of-fit metrics. These outlier spectra may be batch fit again using a secondary spectrum model with more (or less) components than the primary model.

\subsection{Tuning of task parallelization parameters}
For any batch fitting task, there are generally two key parameters to tune to minimize the overall computing time by balancing the number of worker processes ($n_{\mathrm{W}}$) and the number of tasks per worker process ($n_{\mathrm{T}}$), or \textit{chunk size} in this context \cite{Singh2013}. Their trade-off depends on the complexity of the multicomponent spectrum model and the quality of the initial condition provided to the optimizer. We illustrate the trade-off in the example in SM Fig. \ref{fig:taskparallel}. To facilitate the use of our approach, we provide here a list of potential choices for these parameters for reference. From our experience, batch fitting hundreds to thousands of spectra generally requires $n_{\mathrm{W}} > 10$ for performance optimization, although for low-complexity problems (e.g. with very few spectral components), a smaller number of processes will suffice. it's generally. when there are only 2-3 components in the multicomponent spectrum model, $n_{\mathrm{T}}$ should be a few hundreds to maximize the computational efficiency. When there are 4-6 components per spectrum, $n_{\mathrm{T}}$ should be below 100. When there are 8-10 components per spectrum, $n_{\mathrm{T}}$ should be less than 30-40. Beyond 10 components per spectrum, it is advised not to use $n_{\mathrm{T}}$ larger than 10-20. As a rule of thumb, the product of $n_{\mathrm{W}}$ and $n_{\mathrm{T}}$ should be equal or slightly larger than the total number of fitting tasks. We illustrate the trial cases (as purple boxes) in the example in SM Fig. \ref{fig:taskparallel} for tuning the parallelization parameters using the heuristics discussed here.
\\

\printbibliography[heading=subbibliography, title=\large References]

@article{Spurgeon2020,
author = {Spurgeon, Steven R. and Ophus, Colin and Jones, Lewys and Petford-Long, Amanda and Kalinin, Sergei V. and Olszta, Matthew J. and Dunin-Borkowski, Rafal E. and Salmon, Norman and Hattar, Khalid and Yang, Wei-Chang D. and Sharma, Renu and Du, Yingge and Chiaramonti, Ann and Zheng, Haimei and Buck, Edgar C. and Kovarik, Libor and Penn, R. Lee and Li, Dongsheng and Zhang, Xin and Murayama, Mitsuhiro and Taheri, Mitra L.},
doi = {10.1038/s41563-020-00833-z},
issn = {1476-1122},
journal = {Nature Materials},
month = {oct},
title = {{Towards data-driven next-generation transmission electron microscopy}},
url = {http://www.nature.com/articles/s41563-020-00833-z},
year = {2020}
}

@book{Woodruff2016,
address = {Cambridge},
author = {Woodruff, Phil},
doi = {10.1017/CBO9781139149716},
edition = {3rd},
isbn = {9781139149716},
publisher = {Cambridge University Press},
title = {{Modern Techniques of Surface Science}},
url = {http://ebooks.cambridge.org/ref/id/CBO9781139149716},
year = {2016}
}

@article{Kusne2020,
author = {Kusne, A. Gilad and Yu, Heshan and Wu, Changming and Zhang, Huairuo and Hattrick-Simpers, Jason and DeCost, Brian and Sarker, Suchismita and Oses, Corey and Toher, Cormac and Curtarolo, Stefano and Davydov, Albert V. and Agarwal, Ritesh and Bendersky, Leonid A. and Li, Mo and Mehta, Apurva and Takeuchi, Ichiro},
doi = {10.1038/s41467-020-19597-w},
issn = {2041-1723},
journal = {Nature Communications},
month = {dec},
number = {1},
pages = {5966},
title = {{On-the-fly closed-loop materials discovery via Bayesian active learning}},
url = {http://www.nature.com/articles/s41467-020-19597-w},
volume = {11},
year = {2020}
}

@article{Hofmann2020,
archivePrefix = {arXiv},
arxivId = {2011.11490},
author = {Hofmann, Philip},
eprint = {2011.11490},
journal = {arXiv},
month = {nov},
pages = {2011.11490},
title = {{In operando devices studied by angle-resolved photoemission spectroscopy}},
url = {http://arxiv.org/abs/2011.11490},
year = {2020}
}

@article{Umehara2019,
author = {Umehara, Mitsutaro and Stein, Helge S. and Guevarra, Dan and Newhouse, Paul F. and Boyd, David A. and Gregoire, John M.},
doi = {10.1038/s41524-019-0172-5},
issn = {2057-3960},
journal = {npj Computational Materials},
month = {dec},
number = {1},
pages = {34},
title = {{Analyzing machine learning models to accelerate generation of fundamental materials insights}},
url = {http://www.nature.com/articles/s41524-019-0172-5},
volume = {5},
year = {2019}
}

@article{Attia2020,
author = {Attia, Peter M. and Grover, Aditya and Jin, Norman and Severson, Kristen A. and Markov, Todor M. and Liao, Yang-Hung and Chen, Michael H. and Cheong, Bryan and Perkins, Nicholas and Yang, Zi and Herring, Patrick K. and Aykol, Muratahan and Harris, Stephen J. and Braatz, Richard D. and Ermon, Stefano and Chueh, William C.},
doi = {10.1038/s41586-020-1994-5},
issn = {0028-0836},
journal = {Nature},
month = {feb},
number = {7795},
pages = {397--402},
title = {{Closed-loop optimization of fast-charging protocols for batteries with machine learning}},
url = {http://www.nature.com/articles/s41586-020-1994-5},
volume = {578},
year = {2020}
}

@article{Blackburn1965,
author = {Blackburn, J. A.},
doi = {10.1021/ac60227a013},
issn = {0003-2700},
journal = {Analytical Chemistry},
month = {jul},
number = {8},
pages = {1000--1003},
title = {{Computer Program for Multicomponent Spectrum Analysis Using Least Squares Method.}},
url = {https://pubs.acs.org/doi/abs/10.1021/ac60227a013},
volume = {37},
year = {1965}
}

@article{Ruhe1980,
author = {Ruhe, Axel and Wedin, Per {\AA}ke},
doi = {10.1137/1022057},
issn = {0036-1445},
journal = {SIAM Review},
month = {jul},
number = {3},
pages = {318--337},
title = {{Algorithms for Separable Nonlinear Least Squares Problems}},
url = {http://epubs.siam.org/doi/10.1137/1022057},
volume = {22},
year = {1980}
}

@article{DeJuan2014,
author = {de Juan, Anna and Jaumot, Joaquim and Tauler, Rom{\`{a}}},
doi = {10.1039/C4AY00571F},
issn = {1759-9660},
journal = {Anal. Methods},
number = {14},
pages = {4964--4976},
title = {{Multivariate Curve Resolution (MCR). Solving the mixture analysis problem}},
url = {http://xlink.rsc.org/?DOI=C4AY00571F},
volume = {6},
year = {2014}
}

@article{Kriesten2008,
author = {Kriesten, E. and Mayer, D. and Alsmeyer, F. and Minnich, C.B. and Greiner, L. and Marquardt, W.},
doi = {10.1016/j.chemolab.2008.05.002},
issn = {01697439},
journal = {Chemometrics and Intelligent Laboratory Systems},
month = {oct},
number = {2},
pages = {108--119},
title = {{Identification of unknown pure component spectra by indirect hard modeling}},
url = {https://linkinghub.elsevier.com/retrieve/pii/S0169743908000804},
volume = {93},
year = {2008}
}

@article{Speranza2019,
author = {Speranza, Giorgio and Canteri, Roberto},
doi = {10.1016/j.softx.2019.100282},
issn = {23527110},
journal = {SoftwareX},
month = {jul},
pages = {100282},
title = {{RxpsG a new open project for Photoelectron and Electron Spectroscopy data processing}},
url = {https://linkinghub.elsevier.com/retrieve/pii/S2352711019300378},
volume = {10},
year = {2019}
}

@article{RenedeCotret2018,
author = {{Ren{\'{e}} de Cotret}, Laurent P. and Otto, Martin R. and Stern, Mark J. and Siwick, Bradley J.},
doi = {10.1186/s40679-018-0060-y},
issn = {2198-0926},
journal = {Advanced Structural and Chemical Imaging},
month = {dec},
number = {1},
pages = {11},
title = {{An open-source software ecosystem for the interactive exploration of ultrafast electron scattering data}},
url = {https://ascimaging.springeropen.com/articles/10.1186/s40679-018-0060-y},
volume = {4},
year = {2018}
}

@misc{Xian,
author = {Xian, Rui Patrick},
title = {pesfit},
url = {https://github.com/mpes-kit/pesfit},
howpublished = "\url{https://github.com/mpes-kit/pesfit}"
}

@article{Lammel2008,
author = {L{\"{a}}mmel, Ralf},
doi = {10.1016/j.scico.2007.07.001},
issn = {01676423},
journal = {Science of Computer Programming},
month = {jan},
number = {1},
pages = {1--30},
title = {{Google's MapReduce programming model — Revisited}},
url = {https://linkinghub.elsevier.com/retrieve/pii/S0167642307001281},
volume = {70},
year = {2008}
}

@misc{Newville2019,
author = {Newville, Matt and Otten, Renee and Nelson, Andrew and Ingargiola, Antonino and Stensitzki, Till and Allan, Dan and Fox, Austin and Carter, Faustin and Micha{\l} and Pustakhod, Dima and Ram, Yoav and Glenn and Deil, Christoph and Stuermer and Beelen, Alexandre and Frost, Oliver and Zobrist, Nicholas and Pasquevich, Gustavo and Hansen, Allan L. R. and Spillane, Tim and Caldwell, Shane and Polloreno, Anthony and Andrewhannum and Zimmermann, Julius and Borreguero, Jose and Fraine, Jonathan and Deep-42-thought and Maier, Benjamin F. and Gamari, Ben and Almarza, Anthony},
doi = {10.5281/ZENODO.3588521},
howpublished = {\url{https://doi.org/10.5281/zenodo.3588521}},
journal = {Zenodo},
month = {dec},
title = {lmfit/lmfit-py 1.0.0},
url = {https://doi.org/10.5281/zenodo.3588521{\#}.XqH3g-7jiGU.mendeley},
year = {2019}
}

@manual{DaskDevelopmentTeam2016,
author = {{Dask Development Team}},
title = {{Dask: Library for dynamic task scheduling}},
url = {https://dask.org},
year = {2016}
}

@misc{Oller,
author = {Oller, Sergio},
title = {parmap},
url = {https://github.com/zeehio/parmap},
howpublished = "\url{https://github.com/zeehio/parmap}"
}

@article{Hadjidoukas2020,
author = {Hadjidoukas, P.E. and Bartezzaghi, A. and Scheidegger, F. and Istrate, R. and Bekas, C. and Malossi, A.C.I.},
doi = {10.1016/j.softx.2020.100517},
issn = {23527110},
journal = {SoftwareX},
month = {jul},
pages = {100517},
title = {{torcpy: Supporting task parallelism in Python}},
url = {https://linkinghub.elsevier.com/retrieve/pii/S2352711020300091},
volume = {12},
year = {2020}
}

@InProceedings{ mckinney-proc-scipy-2010,
  author    = { {W}es {M}c{K}inney },
  title     = { {D}ata {S}tructures for {S}tatistical {C}omputing in {P}ython },
  booktitle = { {P}roceedings of the 9th {P}ython in {S}cience {C}onference },
  pages     = { 56 - 61 },
  year      = { 2010 },
  editor    = { {S}t\'efan van der {W}alt and {J}arrod {M}illman },
  doi       = { 10.25080/Majora-92bf1922-00a }
}

@article{Zahn2020,
archivePrefix = {arXiv},
arxivId = {2012.10428},
author = {Zahn, Daniela and Seiler, H{\'{e}}l{\`{e}}ne and Windsor, Yoav William and Ernstorfer, Ralph},
eprint = {2012.10428},
journal = {arXiv},
month = {dec},
pages = {2012.10428},
title = {{Ultrafast lattice dynamics and electron-phonon coupling in laser-excited platinum}},
url = {http://arxiv.org/abs/2012.10428},
year = {2020}
}

@article{Golub2003,
author = {Golub, Gene and Pereyra, Victor},
doi = {10.1088/0266-5611/19/2/201},
issn = {0266-5611},
journal = {Inverse Problems},
month = {apr},
number = {2},
pages = {R1--R26},
title = {{Separable nonlinear least squares: the variable projection method and its applications}},
url = {https://iopscience.iop.org/article/10.1088/0266-5611/19/2/201},
volume = {19},
year = {2003}
}

@article{Grubb2016,
author = {Grubb, Michael P. and Coulter, Philip M. and Marroux, Hugo J. B. and Hornung, Balazs and McMullen, Ryan S. and Orr-Ewing, Andrew J. and Ashfold, Michael N. R.},
doi = {10.1038/nchem.2570},
issn = {1755-4330},
journal = {Nature Chemistry},
month = {nov},
number = {11},
pages = {1042--1046},
title = {{Translational, rotational and vibrational relaxation dynamics of a solute molecule in a non-interacting solvent}},
url = {http://www.nature.com/articles/nchem.2570},
volume = {8},
year = {2016}
}

@article{Pushie2014,
author = {Pushie, M. Jake and Pickering, Ingrid J. and Korbas, Malgorzata and Hackett, Mark J. and George, Graham N.},
doi = {10.1021/cr4007297},
issn = {0009-2665},
journal = {Chemical Reviews},
month = {sep},
number = {17},
pages = {8499--8541},
title = {{Elemental and Chemically Specific X-ray Fluorescence Imaging of Biological Systems}},
url = {https://pubs.acs.org/doi/10.1021/cr4007297},
volume = {114},
year = {2014}
}

@article{Kormanyos2013,
author = {Korm{\'{a}}nyos, Andor and Z{\'{o}}lyomi, Viktor and Drummond, Neil D. and Rakyta, P{\'{e}}ter and Burkard, Guido and Fal'ko, Vladimir I.},
doi = {10.1103/PhysRevB.88.045416},
issn = {1098-0121},
journal = {Physical Review B},
month = {jul},
number = {4},
pages = {045416},
title = {{Monolayer MoS {\textless}math display="inline"{\textgreater} {\textless}msub{\textgreater} {\textless}mrow/{\textgreater} {\textless}mn{\textgreater}2{\textless}/mn{\textgreater} {\textless}/msub{\textgreater} {\textless}/math{\textgreater} : Trigonal warping, the {\textless}math display="inline"{\textgreater} {\textless}mi{\textgreater}$\Gamma${\textless}/mi{\textgreater} {\textless}/math{\textgreater} valley, and spin-orbit coupling effects}},
url = {https://link.aps.org/doi/10.1103/PhysRevB.88.045416},
volume = {88},
year = {2013}
}

@article{Kormanyos2015,
author = {Korm{\'{a}}nyos, Andor and Burkard, Guido and Gmitra, Martin and Fabian, Jaroslav and Z{\'{o}}lyomi, Viktor and Drummond, Neil D and Fal'ko, Vladimir},
doi = {10.1088/2053-1583/2/2/022001},
issn = {2053-1583},
journal = {2D Materials},
month = {apr},
number = {2},
pages = {022001},
title = {k {\textperiodcentered} p theory for two-dimensional transition metal dichalcogenide semiconductors},
url = {http://stacks.iop.org/2053-1583/2/i=2/a=022001?key=crossref.80ccc6cac1fc6ba17eb3283f42586226},
volume = {2},
year = {2015}
}

@article{Yu2014,
author = {Yu, Hongyi and Wu, Yue and Liu, Gui-Bin and Xu, Xiaodong and Yao, Wang},
doi = {10.1103/PhysRevLett.113.156603},
issn = {0031-9007},
journal = {Physical Review Letters},
month = {oct},
number = {15},
pages = {156603},
title = {{Nonlinear Valley and Spin Currents from Fermi Pocket Anisotropy in 2D Crystals}},
url = {https://link.aps.org/doi/10.1103/PhysRevLett.113.156603},
volume = {113},
year = {2014}
}

@article{Shan2015,
author = {Shan, Wen-Yu and Zhou, Jianhui and Xiao, Di},
doi = {10.1103/PhysRevB.91.035402},
issn = {1098-0121},
journal = {Physical Review B},
month = {jan},
number = {3},
pages = {035402},
title = {{Optical generation and detection of pure valley current in monolayer transition-metal dichalcogenides}},
url = {https://link.aps.org/doi/10.1103/PhysRevB.91.035402},
volume = {91},
year = {2015}
}

@article{Kalameitsev2019,
author = {Kalameitsev, A. V. and Kovalev, V. M. and Savenko, I. G.},
doi = {10.1103/PhysRevLett.122.256801},
issn = {0031-9007},
journal = {Physical Review Letters},
month = {jun},
number = {25},
pages = {256801},
title = {{Valley Acoustoelectric Effect}},
url = {https://link.aps.org/doi/10.1103/PhysRevLett.122.256801},
volume = {122},
year = {2019}
}

@article{Lindner2015,
author = {Lindner, Robert R. and Vera-Ciro, Carlos and Murray, Claire E. and Stanimirovi{\'{c}}, Sne{\v{z}}ana and Babler, Brian and Heiles, Carl and Hennebelle, Patrick and Goss, W. M. and Dickey, John},
doi = {10.1088/0004-6256/149/4/138},
issn = {1538-3881},
journal = {The Astronomical Journal},
month = {mar},
number = {4},
pages = {138},
title = {{Autonomous Gaussian decomposition}},
url = {https://iopscience.iop.org/article/10.1088/0004-6256/149/4/138},
volume = {149},
year = {2015}
}

@article{Ament2019,
author = {Ament, Sebastian E. and Stein, Helge S. and Guevarra, Dan and Zhou, Lan and Haber, Joel A. and Boyd, David A. and Umehara, Mitsutaro and Gregoire, John M. and Gomes, Carla P.},
doi = {10.1038/s41524-019-0213-0},
issn = {2057-3960},
journal = {npj Computational Materials},
month = {dec},
number = {1},
pages = {77},
title = {{Multi-component background learning automates signal detection for spectroscopic data}},
url = {http://www.nature.com/articles/s41524-019-0213-0},
volume = {5},
year = {2019}
}

@article{Xian2020,
author = {Xian, R. Patrick and Acremann, Yves and Agustsson, Steinn Y. and Dendzik, Maciej and B{\"{u}}hlmann, Kevin and Curcio, Davide and Kutnyakhov, Dmytro and Pressacco, Federico and Heber, Michael and Dong, Shuo and Pincelli, Tommaso and Demsar, Jure and Wurth, Wilfried and Hofmann, Philip and Wolf, Martin and Scheidgen, Markus and Rettig, Laurenz and Ernstorfer, Ralph},
doi = {10.1038/s41597-020-00769-8},
issn = {2052-4463},
journal = {Scientific Data},
month = {dec},
number = {1},
pages = {442},
title = {{An open-source, end-to-end workflow for multidimensional photoemission spectroscopy}},
url = {http://www.nature.com/articles/s41597-020-00769-8},
volume = {7},
year = {2020}
}

@article{Puppin2019,
author = {Puppin, M. and Deng, Y. and Nicholson, C. W. and Feldl, J. and Schr{\"{o}}ter, N. B. M. and Vita, H. and Kirchmann, P. S. and Monney, C. and Rettig, L. and Wolf, M. and Ernstorfer, R.},
doi = {10.1063/1.5081938},
issn = {0034-6748},
journal = {Review of Scientific Instruments},
month = {2},
number = {2},
pages = {023104},
title = {{Time- and angle-resolved photoemission spectroscopy of solids in the extreme ultraviolet at 500 kHz repetition rate}},
url = {http://aip.scitation.org/doi/10.1063/1.5081938},
volume = {90},
year = {2019}
}

@article{Waldecker2015,
author = {Waldecker, Lutz and Bertoni, Roman and Ernstorfer, Ralph},
doi = {10.1063/1.4906786},
issn = {0021-8979},
journal = {Journal of Applied Physics},
month = {jan},
number = {4},
pages = {044903},
title = {{Compact femtosecond electron diffractometer with 100 keV electron bunches approaching the single-electron pulse duration limit}},
url = {http://aip.scitation.org/doi/10.1063/1.4906786},
volume = {117},
year = {2015}
}

@article{Transtrum2015,
author = {Transtrum, Mark K. and Machta, Benjamin B. and Brown, Kevin S. and Daniels, Bryan C. and Myers, Christopher R. and Sethna, James P.},
doi = {10.1063/1.4923066},
issn = {0021-9606},
journal = {The Journal of Chemical Physics},
month = {jul},
number = {1},
pages = {010901},
title = {{Perspective: Sloppiness and emergent theories in physics, biology, and beyond}},
url = {http://aip.scitation.org/doi/10.1063/1.4923066},
volume = {143},
year = {2015}
}

@article{Nakayama2020,
author = {Nakayama, Yasuo and Kera, Satoshi and Ueno, Nobuo},
doi = {10.1039/D0TC00891E},
issn = {2050-7526},
journal = {Journal of Materials Chemistry C},
number = {27},
pages = {9090--9132},
title = {{Photoelectron spectroscopy on single crystals of organic semiconductors: experimental electronic band structure for optoelectronic properties}},
url = {http://xlink.rsc.org/?DOI=D0TC00891E},
volume = {8},
year = {2020}
}

@article{Alfeld1984,
author = {Alfeld, Peter},
doi = {10.1016/0167-8396(84)90029-3},
issn = {01678396},
journal = {Computer Aided Geometric Design},
month = {nov},
number = {2},
pages = {169--181},
title = {{A trivariate clough—tocher scheme for tetrahedral data}},
url = {https://linkinghub.elsevier.com/retrieve/pii/0167839684900293},
volume = {1},
year = {1984}
}

@article{Virtanen2020,
author = {Virtanen, Pauli and Gommers, Ralf and Oliphant, Travis E. and Haberland, Matt and Reddy, Tyler and Cournapeau, David and Burovski, Evgeni and Peterson, Pearu and Weckesser, Warren and Bright, Jonathan and van der Walt, St{\'{e}}fan J. and Brett, Matthew and Wilson, Joshua and Millman, K. Jarrod and Mayorov, Nikolay and Nelson, Andrew R. J. and Jones, Eric and Kern, Robert and Larson, Eric and Carey, C J and Polat, İlhan and Feng, Yu and Moore, Eric W. and VanderPlas, Jake and Laxalde, Denis and Perktold, Josef and Cimrman, Robert and Henriksen, Ian and Quintero, E. A. and Harris, Charles R. and Archibald, Anne M. and Ribeiro, Ant{\^{o}}nio H. and Pedregosa, Fabian and van Mulbregt, Paul},
doi = {10.1038/s41592-019-0686-2},
issn = {1548-7091},
journal = {Nature Methods},
month = {mar},
number = {3},
pages = {261--272},
title = {{SciPy 1.0: fundamental algorithms for scientific computing in Python}},
url = {http://www.nature.com/articles/s41592-019-0686-2},
volume = {17},
year = {2020}
}

@article{Singh2017,
author = {Singh, Sanjay and Dutta, B. and D'Souza, S. W. and Zavareh, M. G. and Devi, P. and Gibbs, A. S. and Hickel, T. and Chadov, S. and Felser, C. and Pandey, D.},
doi = {10.1038/s41467-017-00883-z},
issn = {2041-1723},
journal = {Nature Communications},
month = {dec},
number = {1},
pages = {1006},
title = {{Robust Bain distortion in the premartensite phase of a platinum-substituted Ni2MnGa magnetic shape memory alloy}},
url = {http://www.nature.com/articles/s41467-017-00883-z},
volume = {8},
year = {2017}
}

@article{Singh2013,
author = {Singh, Navtej and Browne, Lisa-Marie and Butler, Ray},
doi = {10.1016/j.ascom.2013.04.002},
issn = {22131337},
journal = {Astronomy and Computing},
month = {aug},
pages = {1--10},
title = {{Parallel astronomical data processing with Python: Recipes for multicore machines}},
url = {https://linkinghub.elsevier.com/retrieve/pii/S2213133713000085},
volume = {2},
year = {2013}
}

\end{document}